\documentclass[article]{jss}

\usepackage[utf8]{inputenc} 
\usepackage[T1]{fontenc}    
\usepackage{url}            
\usepackage{booktabs}       
\usepackage{amsfonts}       
\usepackage{amsmath}
\usepackage{nicefrac}       
\usepackage{microtype}      
\usepackage{lipsum}
\usepackage{graphicx}
\usepackage{multirow}
\usepackage{rotating}
\usepackage{makecell}
\usepackage{xcolor}

\newcommand{\g}{g(x,y)} 
\newcommand{\gp}{g(r,\theta)} 
\newcommand{\s}{\sigma} 
\newcommand{\sx}{\sigma_x} 
\newcommand{\sy}{\sigma_y}
\newcommand{\tsx}{\tilde{\sigma}_x} 
\newcommand{\tsy}{\tilde{\sigma}_y} 
\newcommand{\p}{\rho} 
\newcommand{\m}{\mu}    
\newcommand{\mx}{\mu_x} 
\newcommand{\my}{\mu_y}
\newcommand{\tmx}{\tilde{\mu}_x} 
\newcommand{\tmy}{\tilde{\mu}_y}
\newcommand{\gt}{p(\theta)} 
\newcommand{\gr}{p(r)} 

\newcommand\bfrac[2]{\frac{\displaystyle #1}{\displaystyle #2}}
\newcommand{\pdr}{\partial}
\newcommand{\blue}[1]{\textcolor{blue}{#1}}

\author{
  Emily A. Cooper \\ 
 University of California, Berkeley \\ Berkeley, CA, USA \\ 
  \And
  Hany Farid \\
    University of California, Berkeley \\ Berkeley, CA, USA
}
\title{A Toolbox for the Radial and Angular Marginalization of Bivariate Normal Distributions}

\Plainauthor{Emily A. Cooper, Hany Farid} 
\Plaintitle{A Toolbox for the Radial and Angular Marginalization of Bivariate Normal Distributions} 
\Shorttitle{Radial and Angular Marginalization of Normals} 

\Abstract{
  Bivariate normal distributions are often used to describe the joint probability density of a pair of random variables. These distributions arise across many domains, from telecommunications, to meteorology, ballistics, and computational neuroscience. In these applications, it is often useful to radially and angularly marginalize (i.e.,~under a polar transformation) the joint probability distribution relative to the coordinate system's origin. This marginalization is trivial for a zero-mean, isotropic distribution, but is non-trivial for the most general case of a non-zero-mean, anisotropic distribution with a non-diagonal covariance matrix. Across domains, a range of solutions with varying degrees of generality have been derived. Here, we provide a concise summary of analytic solutions for the polar marginalization of bivariate normal distributions. This report accompanies a Matlab (Mathworks, Inc.) and R toolbox that provides closed-form and numeric implementations for the marginalizations described herein.
}

\Plainkeywords{bivariate normal distribution, two-dimensional Gaussian distribution, polar marginalization, integral over an offset circle, circular and directional distributions, MATLAB, R} 

\Address{
  Emily A. Cooper \\ 
  School of Optometry \\ Helen Wills Neuroscience Institute \\ University of California, Berkeley \\ Berkeley, CA, USA \\ \texttt{emilycooper@berkeley.edu} \\

  Hany Farid \\
  Electrical Engineering \& Computer Sciences \\ School of Information \\  University of California, Berkeley \\ Berkeley, CA, USA \\ \texttt{hfarid@berkeley.edu}
}

\begin{document}


\section{Introduction}

\begin{figure}[t]
    \centering
    \includegraphics[width=0.8\linewidth]{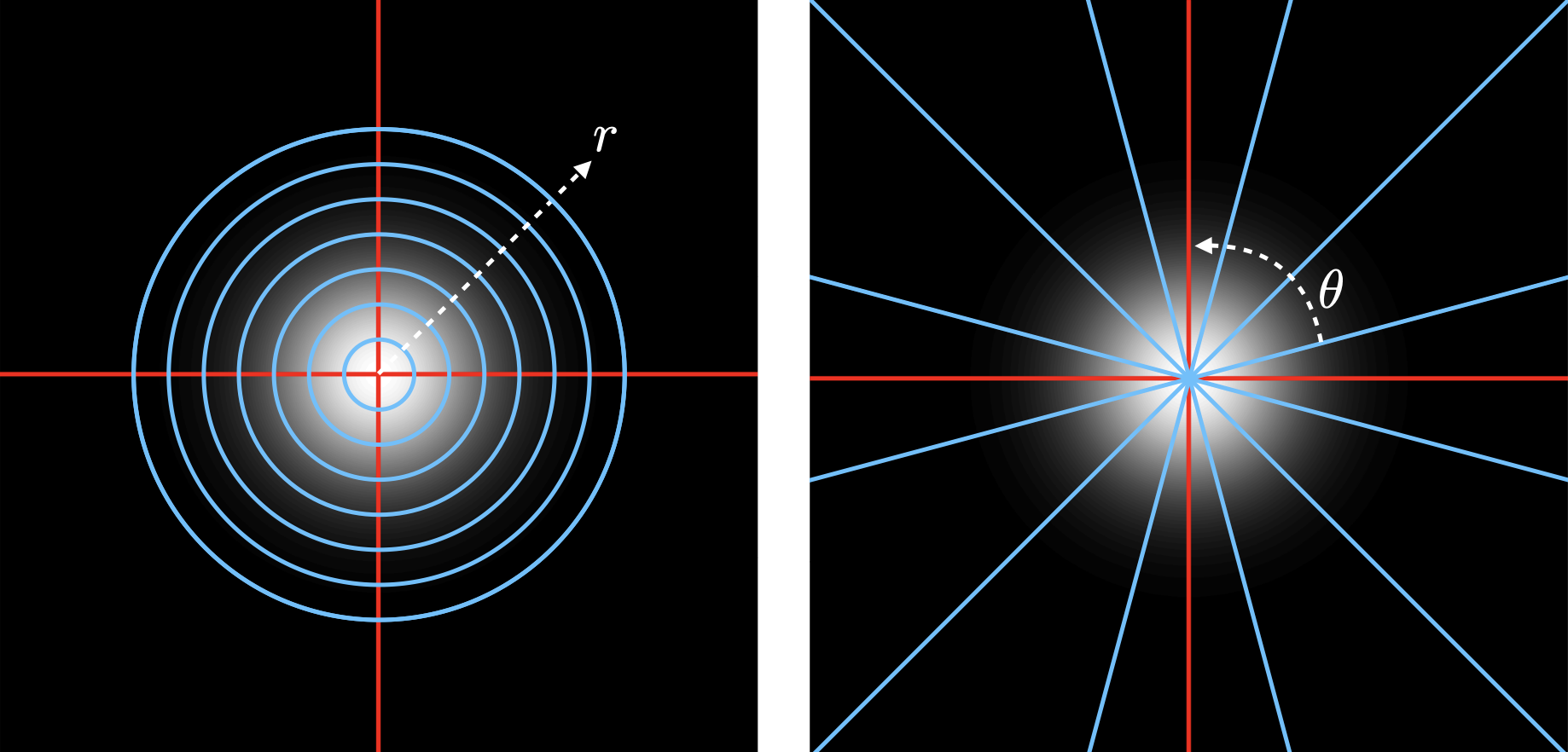} 
    \caption{Angular marginalization (left) in which the probability of a bivariate normal distribution is specified in terms of distance ($r$) from the origin, independent of orientation; and radial  marginalization (right) in which the probability is specified in terms of orientation ($\theta$), independent of distance from the origin.}
    \label{fig:marginilization}
\end{figure}

The need to compute polar marginalizations of bivariate normal probability distributions (i.e.,~two-dimensional Gaussians) arises in many disciplines. Polar marginalization refers to marginalizing the $2$-D distribution to yield $1$-D distributions over radius or angle. {\em Angular marginalization} yields the probability as a function of distance from a specified origin (Figure~\ref{fig:marginilization}, left). This calculation can be performed by integrating over angle in polar coordinates. {\em Radial marginalization} yields the probability as a function of angular direction from a specified origin (Figure~\ref{fig:marginilization}, right). This calculation can be performed by integrating over radius in polar coordinates. 

While analytic solutions exist for these calculations, the documentation for these solutions is spread across the scientific literature of disparate fields, and to our knowledge there exists no single resource that brings these solutions together in one place. To address this gap, we have created the RAMBiNo Toolbox (Radial and Angular Marginalization of Bivariate Normals) to provide a flexible and easy-to-use resource for efficiently computing, in Matlab and R, analytic and numeric polar marginalizations of bivariate normal distributions. The toolbox is available at \url{https://github.com/eacooper/RAMBiNo}. This accompanying report provides the mathematical foundations for these solutions, including direct derivations for the most common calculations. Before presenting these solutions, we first provide a brief background on the applications of polar marginalization and previous work providing exact and approximate solutions.  

The topic of angular marginalization of bivariate normal distributions --- specifying the $1$-D distribution as a function of radius --- has notably arisen in telecommunications and ballistics analysis (e.g.,~\cite{rice1944,rice1945,weil1954distribution,chew1962distribution,gilliland1962integral,grubbs,gillilandandhansen}). The simplest case occurs when the bivariate distribution is centered at the origin (i.e.,~zero-mean and isotropic). In the 1800's, Lord Rayleigh described a closed-form solution to this problem, arising during the analysis of distributions of vibration amplitudes~\cite{rayleigh}. This so-called {\em Rayleigh distribution}, however, has limited practical application to other real-world data. More general analytic (but not closed-form) solutions have subsequently been described for either offset (non-zero-mean) or anisotropic distributions (e.g.,~\cite{rice1944,rice1945,gilliland1962integral,chew1962distribution}). Notably, the {\em Rice} or {\em Rician distribution} (named for Stephen Rice) describes the angular marginalization for a non-zero-mean, isotropic distribution. These more general analytic solutions rely on a modified Bessel function and thus require an approximation to this infinite series~\cite{bowman2012introduction}. Such approximations, however, are typically found in most modern scientific computing environments, and are easy to implement. The most general cases of non-zero-mean and anisotropic bivariate normals (with and without diagonal covariance matrices) also require various infinite series approximations, for which efficient approximations with minimal error have been discussed (e.g.,~\cite{weil1954distribution,gilliland1962integral,ruben1962,gillilandandhansen}). 

On the other hand, the topic of radial marginalization of bivariate normal distributions --- specifying the $1$-D distribution as a function of angle --- has long been of interest in meteorology with respect to calculating wind directions and has also arisen in ecology and neuroscience (e.g.,~\cite{brooks,scott1956regression,mcwilliams,crutcher1962,kendall,carta2009,rokers}). An analytic solution to the most general case of this problem, in which the distribution is offset from the origin (non-zero-mean), anisotropic, and with a non-diagonal covariance matrix, has been widely established~\cite{mardia1972}. In addition to this general solution, more specialized radial marginalizations are useful for their simplicity and computational efficiency. 

\begin{figure}[p]
\begin{center}
\includegraphics[]{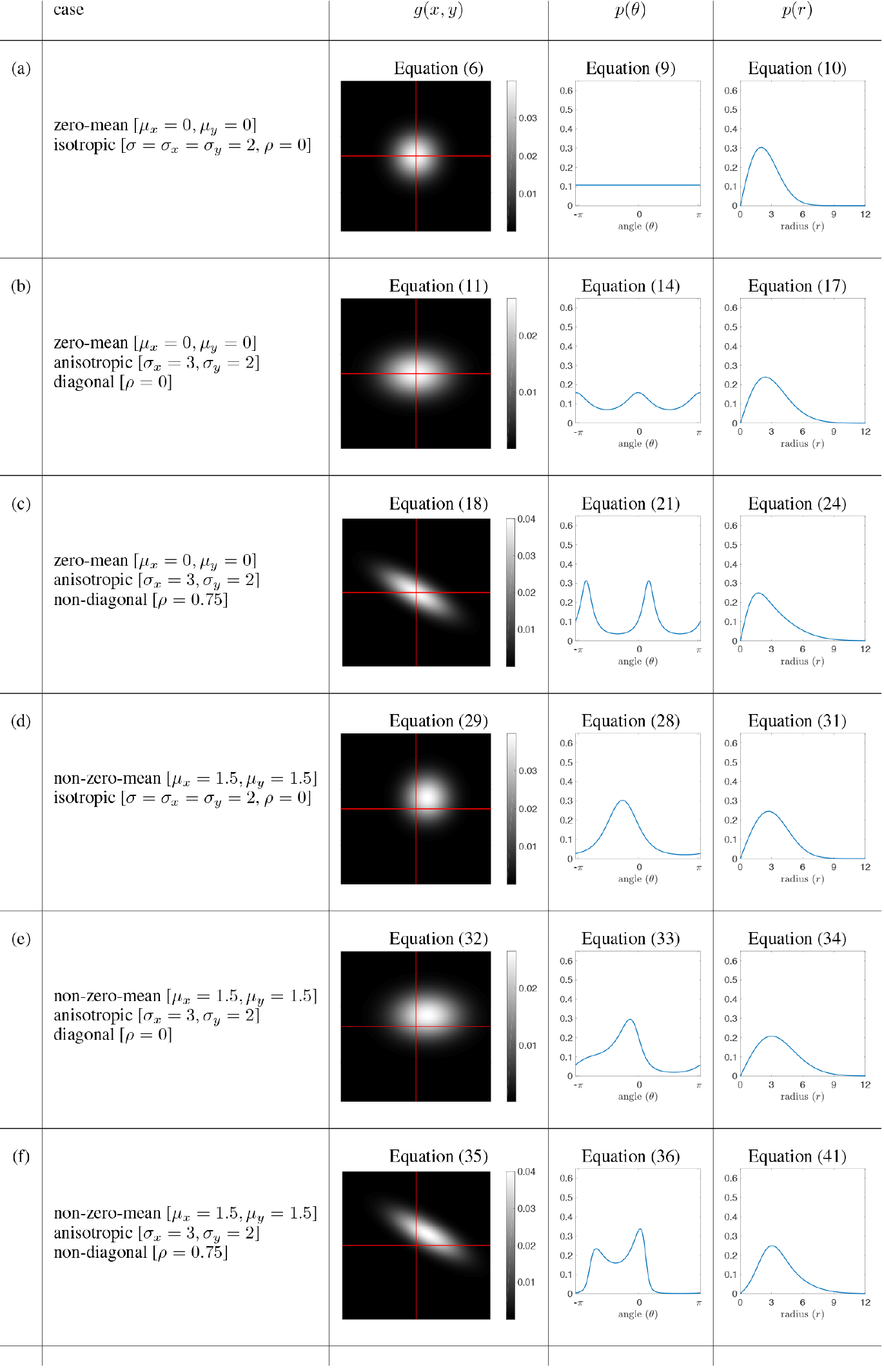}
\end{center}
\label{fig:all}
\caption{The angular and radial marginalization of the bivariate normal distribution (the origin is at the center of the horizontal/vertical red axes).}
\end{figure}

In the following sections, we present the closed-form and analytic solutions for marginalizing bivariate normal distributions in polar coordinates, starting with the specific case of a zero-mean, isotropic distribution and increasing in generality from there. A concise summary of these results are presented in Table~\ref{tab:all} and Figure~\ref{fig:all}, and Matlab code snippets for evaluating these marginalizations may be found in Section~\ref{sec:the-toolbox}. See~\cite{kobayashi,Jammal2001,mardia1972,mardia2009directional} for background on related mathematical problems and distributions.

\noindent\textbf{Preliminaries:} A bivariate normal distribution in spatial parameters $x$ and $y$ and with mean $\mx, \my$, variance $\sx, \sy$, and covariance, $\p$ is defined to be:
\begin{eqnarray}
    \g & = & \bfrac{1}{2\pi\sqrt{|\Sigma|}} \exp \left( -\tfrac{1}{2} \left( \vec{x}-\vec{\m})^T \Sigma^{-1} (\vec{x}-\vec{\m} \right) \right),
\end{eqnarray}
where $\vec{x}^T = \begin{pmatrix}x & y\end{pmatrix}$, $\vec{\m}^T = \begin{pmatrix}\mx & \my\end{pmatrix}$, and where the covariance matrix $\Sigma$, its inverse, and determinant are:
\begin{eqnarray}
    \Sigma ~=~ \begin{pmatrix} \sx^2 & \p\sx\sy \cr \p\sx\sy & \sy^2 \end{pmatrix} \qquad
        \Sigma^{-1} ~=~ \bfrac{1}{|\Sigma|} \begin{pmatrix} \sy^2 & -\p\sx\sy \cr -\p\sx\sy & \sx^2 \end{pmatrix} \qquad
    |\Sigma| ~=~ \sx^2\sy^2(1-\p^2).
\label{eqn:covariance}
\end{eqnarray}
Because we will consider multiple variations of this general distribution (zero vs. non-zero mean, isotropic vs. non-isotropic, and diagonal vs. non-diagonal covariance matrix), we will rewrite the above $2$-D bivariate normal distribution by expanding the matrix/vector notation to yield:
\begin{scriptsize}
\begin{eqnarray}
    \g & = & \bfrac{1}{2\pi\sx\sy\sqrt{1-\p^2}} \exp \left( -\bfrac{1}{2(1-\p^2)} \left( \bfrac{(x-\mx)^2}{\sx^2} + \bfrac{(y-\my)^2}{\sy^2} - \bfrac{2\p(x-\mx)(y-\my)}{\sx\sy} \right) \right).
\label{eqn:normal}
\end{eqnarray}
\end{scriptsize}
In order to marginalize this normal distribution, and its variants, over the distance from the origin (radius $r$) and orientation (angle $\theta$), we will first convert from rectangular to polar coordinates by substituting $x = r\cos(\theta)$ and $y = r\sin(\theta)$ to yield $\gp$. Because we will be considering the integral of this normal distribution under a variable substitution, the double integral takes the form:
\begin{eqnarray}
    \int_r \int_\theta  |J| \gp dr d\theta,
\end{eqnarray}
where $|J|$ denotes the determinant of the Jacobian of the rectangular to polar variable substitution and is defined to be:
\begin{eqnarray}
    |J| \quad = \quad
    \left| \begin{matrix} \bfrac{\pdr x}{\pdr r}  & \bfrac{\pdr x}{\pdr \theta} \cr \bfrac{\pdr y}{\pdr r} & \bfrac{\pdr y}{\pdr \theta} \end{matrix} \right| \quad = \quad
    \left| \begin{matrix} \cos(\theta) & -r\sin(\theta) \cr \sin(\theta) & r\cos(\theta) \end{matrix} \right| \quad = \quad
    r.
\label{eqn:jacobian}
\end{eqnarray}
%
%

\noindent \textbf{Notation:} Throughout this paper, we adopt  the following notation:
\begin{itemize} \itemsep -0.1cm
\item $\g$: bivariate normal distribution in rectangular coordinates
\item $\gp$: bivariate normal distribution expressed in polar coordinates
\item $\gt$: marginal distribution of a bivariate normal marginalized over radius
\item $\gr$: marginal distribution of a bivariate normal marginalized over angle
\item $e^{x} = \exp(x)$: exponential
\item $\phi(x)$: normal probability density function (pdf) with zero-mean and unit-variance
\item $\Phi(x)$: normal cumulative distribution function (cdf) with zero-mean and unit-variance 
\item $I_n(x)$: $n^{th}$-order modified Bessel function of the first kind\footnote{We assume that even-order, modified Bessel functions of the first kind are symmetric, and odd-order are asymmetric, consistent with Matlab's {\tt besseli} function.}
\item $\Gamma(x) = (x-1)!$: gamma function
\item $|A|$: determinant of matrix $A$
\end{itemize}

\section{Marginalization}

In each of the following six sections, we consider the polar marginalization of increasingly more general versions of the bivariate normal distribution with zero- and non-zero-mean, isotropic and non-isotropic, and diagonal and non-diagonal covariance matrix.

\subsection{Zero-Mean, Isotropic}

A zero-mean ($\mx=\my=0$), isotropic ($\s=\sigma_x=\sigma_y$ and $\p=0$) bivariate normal distribution, Equation~(\ref{eqn:normal}), is given by:
\begin{eqnarray}
    \g & = & \bfrac{1}{2\pi\s^2} \exp \left( -\bfrac{x^2 + y^2}{2\s^2} \right).
\label{eqn:a-g}
\end{eqnarray}
In order to marginalize this distribution over the distance ($r$) from the origin and orientation ($\theta$), we first convert from rectangular to polar coordinates by substituting $x = r\cos(\theta)$ and $y = r\sin(\theta)$:
\begin{eqnarray}
    \gp & = & \bfrac{1}{2\pi\s^2} \exp\left( -\bfrac{(r\cos(\theta))^2 + (r\sin(\theta))^2}{2\s^2} \right) \nonumber \\
        & = & \bfrac{1}{2\pi\s^2}\exp \left( -\bfrac{r^2}{2\s^2} \right).
\label{eqn:a-gp}
\end{eqnarray}

\vspace{1em}
\noindent\textbf{Probability as a function of angle:} The marginalization over the radius $r$ is defined to be:
\begin{eqnarray}
    \gt & = & \int_0^\infty \bfrac{r}{2\pi\s^2}\exp \left( -\bfrac{r^2}{2\s^2} \right) dr,
\end{eqnarray}
where the additional multiplicative term $r$ is the determinant of the Jacobian $|J|$, Equation~(\ref{eqn:jacobian}). Substituting $\tilde{r} = -\tfrac{r^2}{2\s^2}$ and $\tfrac{d\tilde{r}}{dr} = \tfrac{-r}{\s^2}$, and reversing the integral bounds to absorb the negation in the variable substitution, yields:
\begin{eqnarray}
    \gt  & = & \bfrac{1}{2\pi} \int_{-\infty}^0 \exp(\tilde{r}) d\tilde{r} \nonumber \\
         & = & \bfrac{1}{2\pi} (\exp(0) - \exp(-\infty)) \nonumber \\
         & = & \bfrac{1}{2\pi}.
\label{eqn:a-gt}
\end{eqnarray}

\vspace{1em}
\noindent\textbf{Probability as a function of radius:} Because the polar representation of this normal distribution, Equation~(\ref{eqn:a-gp}), is not a function of angle $\theta$, the marginalization over angle is trivial, yielding the {\em Rayleigh} distribution:
\begin{eqnarray}
    \gr  & = & \int_0^{2\pi} \bfrac{r}{2\pi\s^2}\exp \left( -\bfrac{r^2}{2\s^2} \right) d\theta \nonumber \\
        & = &  2\pi \left( \bfrac{r}{2\pi\s^2}\exp \left( -\bfrac{r^2}{2\s^2} \right) \right) - 0 \left(  \bfrac{r}{2\pi\s^2}\exp \left( -\bfrac{r^2}{2\s^2} \right) \right)\nonumber \\
        & = & \bfrac{r}{\s^2}\exp \left( -\bfrac{r^2}{2\s^2} \right).
\label{eqn:a-gr}
\end{eqnarray}
%
%
\subsection{Zero-Mean, Anisotropic, Diagonal Covariance}

A zero-mean ($\mx=\my=0$), anisotropic ($\sigma_x \neq \sigma_y$) bivariate normal distribution with a diagonal covariance matrix ($\p=0$), Equation~(\ref{eqn:normal}), is given by:
\begin{eqnarray}
    \g & = & \bfrac{1}{2\pi\sx\sy} \exp\left( -\left(\bfrac{x^2}{2\sx^2} + \bfrac{y^2}{2\sy^2} \right) \right).
\label{eqn:b-g}
\end{eqnarray}

\vspace{1em}
\noindent\textbf{Probability as a function of angle:} In order to marginalize this distribution over the distance ($r$) from the origin, we first convert from rectangular to polar coordinates by substituting $x = r\cos(\theta)$ and $y = r\sin(\theta)$:
\begin{eqnarray}
    \gp & = & \bfrac{1}{2\pi\sx\sy} \exp\left( -\left(\bfrac{(r\cos(\theta))^2}{2\sx^2} + \bfrac{(r\sin(\theta))^2}{2\sy^2} \right) \right) \nonumber \\
    & = & \bfrac{1}{2\pi\sx\sy} \exp\left( -r^2\left(\bfrac{\cos^2(\theta)}{2\sx^2} + \bfrac{\sin^2(\theta)}{2\sy^2} \right) \right) \nonumber \\
    & = & \bfrac{1}{a} \exp(-br^2),
\end{eqnarray}
where $a=2\pi\sx\sy$ and $b = \tfrac{\cos^2(\theta)}{2\sx^2} + \tfrac{\sin^2(\theta)}{2\sy^2}$.  The marginalization over the radius $r$ is then defined to be:
\begin{eqnarray}
    \gt \quad = \quad \int_0^{\infty} \bfrac{r}{a} \exp(-br^2) dr  \quad = \quad \bfrac{1}{a} \int_0^{\infty}  r \exp(-br^2) dr,
\end{eqnarray}
where the additional multiplicative term $r$ is the determinant of the Jacobian $|J|$, Equation~(\ref{eqn:jacobian}). This integral is of the general form $\int_0^\infty r^m \exp(-\alpha r^n) dr$, where in our case $m=1$, $n=2$, and $\alpha=b$. The closed-form solution of this integral is $\tfrac{1}{n} \alpha^{-\tfrac{m+1}{n}} \Gamma\left(\tfrac{m+1}{n}\right)$, which in our case yields:
\begin{eqnarray}
    \gt \quad = \quad \bfrac{1}{a}\left( \bfrac{1}{2}b^{-1} \Gamma(1) \right) \quad = \quad \bfrac{1}{2ab},
\label{eqn:b-gt}
\end{eqnarray}
where  $a=2\pi\sx\sy$ and $b = \tfrac{\cos^2(\theta)}{2\sx^2} + \tfrac{\sin^2(\theta)}{2\sy^2}$. Note that for an isotropic normal, Equation~(\ref{eqn:a-g}), where $\s=\sx=\sy$, then $a={2\pi\s^2}$, $b=\tfrac{1}{2\s^2}$, and $\gt$ reduces to $\tfrac{1}{2\pi}$, as in Equation~(\ref{eqn:a-gt}). 

\vspace{1em}
\noindent\textbf{Probability as a function of radius:}  In order to marginalize over angle $\theta$, the bivariate normal is again converted from rectangular to polar coordinates:
\begin{eqnarray}
    \gp & = & \bfrac{1}{2\pi\sx\sy} \exp\left( -\left(\bfrac{(r\cos(\theta))^2}{2\sx^2} + \bfrac{(r\sin(\theta))^2}{2\sy^2} \right) \right) \nonumber \\
        & = & \bfrac{1}{2\pi\sx\sy} \exp\left( -r^2 \left( \bfrac{ \tfrac{1}{2} (1+\cos(2\theta))}{2\sx^2} +
              \bfrac{\tfrac{1}{2} (1-\cos(2\theta))}{2\sy^2} \right) \right) \nonumber \\
        & = & \bfrac{1}{2\pi\sx\sy} 
                \exp\left( -r^2 \left(\bfrac{1}{(2\sx)^2} + \bfrac{1}{(2\sy)^2}\right) \right)
                \exp\left( -r^2 \left(\bfrac{\cos(2\theta)}{(2\sx)^2} - \bfrac{\cos(2\theta)}{(2\sy)^2}\right) \right) \nonumber \\
        & = &  \bfrac{1}{2\pi\sx\sy} \exp(-a r^2) \exp( -b r^2\cos(2\theta) ),
\end{eqnarray}
where $a = \frac{\sx^2 + \sy^2}{(2\sx\sy)^2}$ and $b = \frac{\sx^2 - \sy^2}{(2\sx\sy)^2}$.
The marginalization over the angle $\theta$ is then defined to be:
\begin{eqnarray}
    \gr & = & \int_0^{2\pi} \bfrac{r}{2\pi\sx\sy} \exp(-a r^2) \exp( -b r^2\cos(2\theta) ) d\theta \nonumber \\
        & = & \bfrac{r}{2\pi\sx\sy} \exp(-a r^2) \int_0^{2\pi}  \exp( -b r^2\cos(2\theta) ) d\theta.
\end{eqnarray}
The zeroth-order modified Bessel function of the first kind is defined to be \linebreak $I_0(\alpha) = 2\pi \int_{0}^{2\pi} \exp(\alpha \cos(\theta)) d\theta$, and so the above integral, after a variable substitution of $\tilde{\theta}=2\theta$ and $\tfrac{d\tilde{\theta}}{d\theta} = 2$, is given by:
\begin{eqnarray}
    \gr & = & \bfrac{r}{\sx\sy} \exp\left(-a r^2\right) I_0(-b r^2),
\label{eqn:b-gr}
\end{eqnarray}
where the constants are $a = \frac{\sx^2 + \sy^2}{(2\sx\sy)^2}$ and $b = \frac{\sx^2 - \sy^2}{(2\sx\sy)^2}$. Note that $\gr$ is the same whether $\sx > \sy$ or $\sx < \sy$. Also note that for an isotropic normal, Equation~(\ref{eqn:a-g}), where $\s=\sx=\sy$, then $a=\tfrac{1}{2\s^2}$, $b=0$, $I_0(0)=1$, and $\gr$ reduces to $\tfrac{r}{\s^2} \exp\left(\tfrac{-r^2}{2\s^2}\right)$, as in Equation~(\ref{eqn:a-gr}).

\subsection{Zero-Mean, Anisotropic, Non-Diagonal Covariance}
\label{sec:c}

A zero-mean ($\mx=\my=0$), anisotropic ($\sigma_x \neq \sigma_y$) bivariate normal distribution, with a non-diagonal covariance matrix ($\p \neq 0$), is given by:
\begin{eqnarray}
    \g & = & \bfrac{1}{2\pi\sx\sy\sqrt{1-\p^2}} \exp\left(-\bfrac{1}{2(1-\p^2)}\left( \bfrac{x^2}{\sx^2} + \bfrac{y^2}{\sy^2} - \bfrac{2\p xy}{\sx\sy}\right) \right).
\label{eqn:c-g}
\end{eqnarray}

\vspace{1em}
\noindent\textbf{Probability as a function of angle:} In order to marginalize this distribution over the distance ($r$) from the origin, we first convert from rectangular to polar coordinates by substituting $x = r\cos(\theta)$ and $y = r\sin(\theta)$:
\begin{scriptsize}
\begin{eqnarray}
    \gp & = &  \bfrac{1}{2\pi\sx\sy\sqrt{1-\p^2}} \exp\left(-\bfrac{1}{2(1-\p^2)}\left( \bfrac{(r\cos(\theta))^2}{\sx^2} + \bfrac{(r\sin(\theta))^2}{\sy^2} - \bfrac{2\p (r\cos(\theta))(r\sin(\theta))}{\sx\sy}\right) \right) \nonumber \\
        & = & \bfrac{1}{2\pi\sx\sy\sqrt{1-\p^2}} \exp\left(-r^2 \bfrac{1}{2(1-\p^2)}\left( \bfrac{\cos^2(\theta)}{\sx^2} + \bfrac{\sin^2(\theta)}{\sy^2} - \bfrac{2\p \cos(\theta)\sin(\theta)}{\sx\sy}\right) \right) \nonumber \\
        & = & \bfrac{1}{a} \exp(-br^2),
\end{eqnarray}
\end{scriptsize}
where $a = 2\pi\sx\sy\sqrt{1-\p^2}$ and $b = \tfrac{1}{2(1-\p^2)}\left( \tfrac{\cos^2(\theta)}{\sx^2} + \tfrac{\sin^2(\theta)}{\sy^2} - \tfrac{2\p \cos(\theta)\sin(\theta)}{\sx\sy}\right)$.  The marginalization over the radius $r$ is then defined to be:
\begin{eqnarray}
    \gt \quad = \quad \int_0^{\infty} \bfrac{r}{a} \exp(-br^2) dr \quad = \quad \bfrac{1}{a} \int_0^{\infty} r \exp(-br^2) dr.
\end{eqnarray}
where the additional multiplicative term $r$ is the determinant of the Jacobian $|J|$, Equation~(\ref{eqn:jacobian}). This integral is of the general form $\int_0^\infty r^m \exp(-\alpha r^n) dr$, where in our case $m=1$, $n=2$, and $\alpha=b$. The closed-form solution of this integral is $\tfrac{1}{n} \alpha^{-\tfrac{m+1}{n}} \Gamma\left(\tfrac{m+1}{n}\right)$, which in our case yields:
\begin{eqnarray}
    \gt \quad = \quad \bfrac{1}{a}\left( \bfrac{1}{2}b^{-1} \Gamma(1) \right) \quad = \quad \bfrac{1}{2ab},
\label{eqn:c-gt}
\end{eqnarray}
where, $a = 2\pi\sx\sy\sqrt{1-\p^2}$ and $b = \tfrac{1}{2(1-\p^2)}\left( \tfrac{\cos^2(\theta)}{\sx^2} + \tfrac{\sin^2(\theta)}{\sy^2} - \tfrac{2\p \cos(\theta)\sin(\theta)}{\sx\sy}\right)$.

\vspace{1em}
\noindent\textbf{Probability as a function of radius:}  In order to marginalize over angle $\theta$, we consider a rotation, by $\tfrac{1}{2}\tan^{-1}\left( \tfrac{2\p\sx\sy}{\sx^2 - \sy^2} \right)$, of the coordinate system that aligns the bivariate distribution along the $x-$ and $y-$axes. In this new coordinate system, the mean remains at the origin, and the variances $\tsx, \tsy$ are the eigenvalues of the covariance matrix, Equation~(\ref{eqn:covariance}):
\begin{eqnarray}
    \label{eqn:c-tsx}
    \tsx^2 & = & \bfrac{\sx^2+\sy^2}{2} + \sqrt{\bfrac{(\sx^2+\sy^2)^2}{4} - (\sx^2\sy^2 - \p^2\sx^2\sy^2)} \\
    \label{eqn:c-tsy}
    \tsy^2 & = & \bfrac{\sx^2+\sy^2}{2} - \sqrt{\bfrac{(\sx^2+\sy^2)^2}{4} - (\sx^2\sy^2 - \p^2\sx^2\sy^2)}.
\end{eqnarray}
In this rotated coordinate system, the marginalization takes the same form as for a zero-mean, anisotropic normal distribution with a diagonal covariance matrix, Equation~(\ref{eqn:b-gr}):
\begin{eqnarray}
    \gr & = &  \bfrac{r}{\tsx\tsy} \exp\left(-a r^2\right) I_0(-b r^2),
\label{eqn:c-gr}
\end{eqnarray}
where the constants are $a = \frac{\tsx^2 + \tsy^2}{(2\tsx\tsy)^2}$ and $b = \frac{\tsx^2 - \tsy^2}{(2\tsx\tsy)^2}$.

\subsection{Non-zero-mean, Isotropic}

A non-zero-mean ($(\mx,\my) \neq (0,0)$), isotropic ($\s=\sigma_x=\sigma_y$ and $\p=0$) bivariate normal distribution is given by:
\begin{eqnarray}
    \g & = & \bfrac{1}{2\pi\s^2} \exp\left( - \bfrac{(x-\mx)^2 + (y-\my)^2}{2\s^2} \right).
\label{eqn:d-g}
\end{eqnarray}

\vspace{1em}
\noindent\textbf{Probability as a function of angle:} In order to marginalize this distribution over the distance ($r$) from the origin, we first convert from rectangular to polar coordinates by substituting $x = r\cos(\theta)$ and $y = r\sin(\theta)$:
\begin{eqnarray}
    \gp & = & \bfrac{1}{2\pi\s^2} \exp\left( - \bfrac{(r\cos(\theta)-\mx)^2 + (r\sin(\theta)-\my)^2}{2\s^2} \right) \nonumber \\
        & = & \bfrac{1}{2\pi\s^2} \exp\left( - \bfrac{r^2 - 2r(\mx\cos(\theta) + \my\sin(\theta)) + (\mx^2 + \my^2)}{2\s^2}  \right) \nonumber \\
        & = & \bfrac{1}{2\pi\s^2} \exp\left( - \bfrac{\mx^2 + \my^2}{2\s^2}  \right) \exp\left( - \bfrac{r^2 - 2r(\mx\cos(\theta) + \my\sin(\theta))}{2\s^2} \right).
\end{eqnarray}
The marginalization over the radius $r$ is then defined to be:
\begin{scriptsize}
\begin{eqnarray}
    \gt & = & \bfrac{1}{2\pi\s^2} \exp\left( - \bfrac{\mx^2 + \my^2}{2\s^2} \right) \int_0^\infty r  \exp\left( - \bfrac{r^2 - 2r(\mx\cos(\theta) + \my\sin(\theta))}{2\s^2} \right) dr \nonumber \\
        & = & \bfrac{1}{2\pi} \exp\left( - \bfrac{\mx^2 + \my^2}{2\s^2}  \right) \bfrac{1}{\s^2} \int_0^\infty r \exp \left( -\bfrac{1}{2}\bfrac{1}{\s^2} (r^2 - 2r(\mx\cos(\theta) + \my\sin(\theta)) ) \right) dr,
\end{eqnarray}
\end{scriptsize}
where the additional multiplicative term $r$ is the determinant of the Jacobian $|J|$, Equation~(\ref{eqn:jacobian}). This integral is of the general form $\alpha^2 \int_0^\infty r \exp \left( -\tfrac{1}{2} \alpha^2 (r^2 - 2r\beta) \right) dr$, where in our case $\alpha = \tfrac{1}{\s}$ and $\beta = \mx\cos(\theta) + \my\sin(\theta)$. The solution of this integral is $1 + \sqrt{2\pi}\alpha\beta \exp\left(\tfrac{1}{2}\alpha^2\beta^2\right) \Phi(\alpha\beta)$, (see~\cite{mardia1972}, p. 52). In our case, we will combine the $\alpha$ and $\beta$ terms by defining  $a = \tfrac{1}{\s}\sqrt{\mx^2 + \my^2}$ and $b = \tfrac{1}{\s}(\mx\cos(\theta) + \my\sin(\theta))$, yielding:
\begin{eqnarray}
    \gt & = &  \bfrac{1}{2\pi} \exp \left( -\bfrac{a^2}{2} \right)
        \left( 1 + \sqrt{2\pi}b \exp\left(\bfrac{b^2}{2} \right) \Phi(b) \right) \nonumber \\
        & = & \bfrac{1}{\sqrt{2\pi}} \left( \bfrac{1}{\sqrt{2\pi}} \exp \left( -\bfrac{a^2}{2} \right) \right)
         \left( 1 + \bfrac{b \Phi(b)}{\bfrac{1}{\sqrt{2\pi}} \exp\left(\bfrac{-b^2}{2} \right)} \right) \nonumber \\
        & = & \bfrac{1}{\sqrt{2 \pi}} \phi(a)\left(1 + \bfrac{b\Phi(b)}{\phi(b)}\right).
\label{eqn:d-gt}  
\end{eqnarray}
 Note that for a zero-mean isotropic normal, Equation~(\ref{eqn:a-g}), where $\mx=\my=0$, then $a=0$, $\phi(a)=\tfrac{1}{\sqrt{2\pi}}$, $b=0$, and $\gt$ reduces to $\tfrac{1}{2\pi}$, as in Equation~(\ref{eqn:a-gt}).

\vspace{1em}
\noindent\textbf{Probability as a function of radius:} In order to marginalize over angle $\theta$, the bivariate normal is again converted from rectangular to polar coordinates:
\begin{eqnarray}
    \gp & = & \bfrac{1}{2\pi\s^2} \exp\left( - \bfrac{(r\cos(\theta)-\mx)^2 + (r\sin(\theta)-\my)^2}{2\s^2} \right) \nonumber \\
        & = &  \bfrac{1}{2\pi\s^2} \exp\left( - \bfrac{r^2 + a^2}{2\s^2} - \bfrac{r (\mu_x \cos(\theta) + \mu_y \sin(\theta))}{\s^2} \right) \nonumber \\ 
        & = & \bfrac{1}{2\pi\s^2} \exp\left( - \bfrac{r^2 + a^2}{2\s^2} - \bfrac{r a \cos(\theta - \psi)}{\s^2} \right),
\label{eqn:d-r}
\end{eqnarray}
where $a = \sqrt{\mu_x^2 + \mu_y^2}$ and $\psi = \tan^{-1}\left(\tfrac{\my}{\mx}\right)$. The last step in the above equation is derived using the following trigonometric identities: $\cos(\alpha-\beta) = \cos(\alpha)\cos(\beta) + \sin(\alpha)\sin(\beta)$, $\cos(\tan^{-1}(\alpha)) = \tfrac{1}{\sqrt{1+\alpha^2}}$, and $\sin(\tan^{-1}(\alpha)) = \tfrac{\alpha}{\sqrt{1+\alpha^2}}$. The marginalization over the radius $r$ is then defined to be:
\begin{eqnarray}
\gr & = & \int_0^{2\pi} \bfrac{r}{2\pi\s^2} \exp\left( - \bfrac{r^2 + a^2}{2\s^2} - \bfrac{r a \cos(\theta - \psi)}{\s^2} \right) d\theta \nonumber \\
    & = & \bfrac{r}{\s^2} \exp\left( - \bfrac{r^2 + a^2}{2\s^2} \right) \bfrac{1}{2\pi} \int_0^{2\pi}  \exp\left( -\bfrac{r a}{\s^2} \cos(\theta - \psi) \right) d\theta,
\end{eqnarray}
The zeroth-order modified Bessel function of the first kind is defined to be \linebreak $I_0(\alpha) = 2\pi \int_{0}^{2\pi} \exp(\alpha \cos(\theta)) d\theta$, and so the above integral, after a variable substitution of $\tilde{\theta}=\theta-\psi$ and $\frac{d\tilde{\theta}}{d\theta}=1$, yields the following {\em Rician} distribution~\cite{rice1944,rice1945}:
\begin{eqnarray}
    \gr & = & \bfrac{r}{\s^2} \exp\left( -\bfrac{r^2 + a^2}{2\s^2} \right) I_0\left(\bfrac{ra}{\s^2}\right),
\label{eqn:d-gr}
\end{eqnarray}
where $a = \sqrt{\mx^2 + \my^2}$. Note that for a zero-mean isotropic normal, Equation~(\ref{eqn:a-g}), where $\mx=\my=0$, then $a=0$, $I_0(0)=1$, and $\gr$ reduces to $ \tfrac{r}{\s^2} \exp\left( -\tfrac{r^2}{2\s^2} \right)$, as in Equation~(\ref{eqn:a-gr}).

\subsection{Non-zero-mean, Ansotropic, Diagonal Covariance}

A non-zero-mean ($(\mx,\my) \neq (0,0)$), anisotropic ($\sigma_x \neq \sigma_y$) bivariate normal distribution, with a diagonal covariance matrix ($\p=0$), is given by:
\begin{eqnarray}
    \g & = & \bfrac{1}{2\pi\sx\sy} \exp\left(-\left(\bfrac{(x-\mx)^2}{2\sx^2} + \bfrac{(y-\my)^2}{2\sy^2} \right) \right).
\label{eqn:e-g}
\end{eqnarray}
For the sake of brevity, we do not derive the polar marginalizations for this distribution, referring the reader to~\cite{mardia1972} and~\cite{weil1954distribution}.

\vspace{1em}
\noindent\textbf{Probability as a function of angle:} The marginalization over radius is given by: 
\begin{eqnarray}
\gt & = & \bfrac{1}{a \sqrt{2\pi \sx^2 \sy^2}} \phi(b) \left(1 + \bfrac{c\Phi(c)}{\phi(c)}\right),
\label{eqn:e-gt}
\end{eqnarray}
where $a = \tfrac{1}{\sx^2}\cos^2(\theta) + \tfrac{1}{\sy^2}\sin^2(\theta)$, $b = \sqrt{\tfrac{\mx^2}{\sx^2} + \tfrac{\my^2}{\sy^2}}$, and $c = \tfrac{1}{\sqrt{a}} \left( \tfrac{\mx}{\sx^2}\cos(\theta) + \tfrac{\my}{\sy^2}\sin(\theta) \right)$.

\vspace{1em}
\noindent\textbf{Probability as a function of radius:} The marginalization over angle is given by:
\begin{eqnarray}
\gr & = & a r \exp\left( -\bfrac{r^2(\sx^2 + \sy^2)}{4 \sx^2 \sy^2} \right) \left( I_0(br^2) I_0(cr) + 2 \sum_{k=1}^{\infty} I_k(br^2)I_{2k}(cr) \cos(2k \psi) \right),
\label{eqn:e-gr}
\end{eqnarray}
where $a = \tfrac{1}{\sx\sy} \exp \left( -\tfrac{\mx^2\sy^2 + \my^2\sx^2}{2\sx^2\sy^2} \right)$, $b = \tfrac{\sx^2 - \sy^2}{4\sx^2\sy^2}$, $c = \sqrt{ \left(\tfrac{\mx}{\sx^2}\right)^2 +  \left(\tfrac{\my}{\sy^2}\right)^2}$, and $\psi = \tan^{-1} \left( \tfrac{\my \sx^2}{\mx \sy^2} \right)$. Note that this solution, introduced by~\cite{weil1954distribution}, requires an infinite sum. This series, however, can be truncated to reliatively few terms leading to minimal error~\cite{weil1954distribution}.
\subsection{Nonzero-Mean, Anisotropic, Non-Diagonal Covariance}
\label{sec:f}

Returning now to where we began, Equation~(\ref{eqn:normal}), a non-zero-mean ($(\mx,\my) \neq (0,0)$), anisotropic ($\sigma_x \neq \sigma_y$) bivariate normal distribution, with a non-diagonal covariance matrix ($\p \neq 0$), is given by:
\begin{scriptsize}
\begin{eqnarray}
    \g & = & \bfrac{1}{2\pi\sx\sy\sqrt{1-\p^2}} \exp\left(-\bfrac{1}{2(1-\p^2)}\left( \bfrac{(x-\mx)^2}{\sx^2} + \bfrac{(y-\my)^2}{\sy^2} - \bfrac{2\p(x-\mx)(y-\my)}{\sx\sy}\right)\right).
\label{eqn:f-g}
\end{eqnarray}
\end{scriptsize}
For the sake of brevity, we do not fully derive the polar marginalizations for this distribution, referring the reader to~\cite{mardia1972} and~\cite{weil1954distribution}.

\vspace{1em}
\noindent\textbf{Probability as a function of angle:}
\begin{eqnarray}
    \gt & = & \bfrac{1}{a}\left(b + cd\Phi(d)\phi\left( \bfrac{c(\mx\sin(\theta) - \my\cos(\theta))}{\sqrt{a}}\right)\right),
\label{eqn:f-gt}
\end{eqnarray}
where $a=c^2(\sy^2\cos^2(\theta) - \p\sx\sy\sin(2\theta) + \sx^2\sin^2(\theta))$, $b=g(\mx,\my;0,0,\sx,\sy,\p)$, \linebreak $c=\tfrac{1}{\sx\sy\sqrt{1-\p^2}}$, and $d=\tfrac{c^2}{\sqrt{a}} (\mx\sy(\sy\cos(\theta) - \p\sx\sin(\theta))+ \my\sx(\sx\sin(\theta) - \p\sy\cos(\theta)))$, and where  $g(x,y;\mx,\my,\sx,\sy,\p)$ denotes a bivariate normal with mean $\mx,\my$, variance $\sx,\sy$ and covariance $\p$ evaluated at $x,y$.

\vspace{1em}
\noindent\textbf{Probability as a function of radius:} In order to marginalize over angle $\theta$, we consider a rotation, by $\omega=\tfrac{1}{2}\tan^{-1}\left( \tfrac{2\p\sx\sy}{\sx^2 - \sy^2} \right)$,  of the coordinate system that aligns the bivariate distribution along the $x-$ and $y-$axes. In this new coordinate system, the mean $\tmx, \tmy$ of the bivariate distribution is:
\begin{eqnarray}
    \label{eqn:f-tmx}
    \tmx & = & \mx\cos(\omega) + \my\sin(\omega) \\
    \label{eqn:f-tmy}
    \tmy & = &  -\mx\sin(\omega) + \my\cos(\omega),
\end{eqnarray}
and the variances $\tsx, \tsy$ are the eigenvalues of the covariance matrix, Equation~(\ref{eqn:covariance}):
\begin{eqnarray}
    \label{eqn:f-tsx}
    \tsx^2 & = & \bfrac{\sx^2+\sy^2}{2} + \sqrt{\bfrac{(\sx^2+\sy^2)^2}{4} - (\sx^2\sy^2 - \p^2\sx^2\sy^2)} \\
    \label{eqn:f-tsy}
    \tsy^2 & = & \bfrac{\sx^2+\sy^2}{2} - \sqrt{\bfrac{(\sx^2+\sy^2)^2}{4} - (\sx^2\sy^2 - \p^2\sx^2\sy^2)}.
\end{eqnarray}
In this rotated coordinate system, the marginalization takes the same form as for a non-zero-mean, anisotropic normal distribution with a diagonal covariance matrix, Equation~(\ref{eqn:e-gr}):
\begin{eqnarray}
    \gr & = & a r \exp\left( -\bfrac{r^2(\tsx^2 + \tsy^2)}{4 \tsx^2 \tsy^2} \right) \left( I_0(br^2) I_0(cr) + 2 \sum_{k=1}^{\infty} I_k(br^2)I_{2k}(cr) \cos(2k \psi) \right),
\label{eqn:f-gr}
\end{eqnarray}
where $a = \tfrac{1}{\tsx\tsy} \exp \left( -\tfrac{\tmx^2\tsy^2 + \tmy^2\tsx^2}{2\tsx^2\tsy^2} \right)$, $b = \tfrac{\tsx^2 - \tsy^2}{4\tsx^2\tsy^2}$, $c = \sqrt{ \left(\tfrac{\tmx}{\tsx^2}\right)^2 +  \left(\tfrac{\tmy}{\tsy^2}\right)^2}$, and $\psi = \tan^{-1}\left( \tfrac{\tmy \tsx^2}{\tmx \tsy^2} \right)$. Note that this solution, introduced by~\cite{weil1954distribution}, requires an infinite sum. This series, however, can be truncated to reliatively few terms leading to minimal error~\cite{weil1954distribution}.
        
\section{Timing}

We compared the run-time of the Matlab code snippets (Section~\ref{sec:the-toolbox}) for the numeric solution and each of the six analytic solutions described above. Shown in the table below is the relative speed of each numeric and analytic solution as compared to the most general analytic solution described in Section~\ref{sec:f} (Equations~(\ref{eqn:f-gt}) and~(\ref{eqn:f-gr})), where the radial and angular variable are each sampled at $1000$ values. Speeds greater than $1.0$ correspond to faster run-time, and values less than $1.0$ correspond to slower run-times. The analytic solutions are significantly faster to evaluate than the numeric solutions, but even within the analytic solutions, the less general bivariate normal distributions are much faster to marginalize, showing the value of having a range of solutions, from least to most general.

\vspace{1em}
\resizebox{1\textwidth}{!}{
\begin{tabular}{c|ccc|ccc|c}
    \hline
    distribution & zero-mean & zero-mean & zero-mean & non-zero-mean & non-zero-mean & non-zero-mean & numeric \\
                 & isotropic & anisotropic & anisotropic & isotropic & anisotropic & anisotropic & \\
                 &           & diagonal & non-diagonal &           & diagonal & non-diagonal & \\
    \hline
    $\gt$ & $4.7$ & $3.2$ & $2.5$ & $1.3$ & $1.1$ & $1.0$ & $0.017$ \\
    \hline
     $\gr$ & $465.6$ & $114.0$ & $101.7$ & $119.9$ & $1.1$ & $1.0$ & $0.130$ \\
    \hline 
\end{tabular}
}
%
%

\def \eqag   {$\g = \bfrac{1}{2\pi\s^2} \exp\left( -\bfrac{x^2 + y^2}{2\s^2} \right)$}
\def \eqagt  {$\gt = \bfrac{1}{2\pi}$}
\def \eqagr  {$\gr = \bfrac{r}{\s^2} \exp\left(-\bfrac{r^2}{2\s^2} \right)$}

\def \eqbg   {$\g = \bfrac{1}{2\pi\sx\sy} \exp\left( -\left(\bfrac{x^2}{2\sx^2} + \bfrac{y^2}{2\sy^2} \right) \right)$}
\def \eqbgt  {$\gt = \bfrac{1}{2ab}$}
\def \eqbgta {$a = 2\pi\sx\sy$}
\def \eqbgtb {$b = \bfrac{\cos^2(\theta)}{2\sx^2} + \bfrac{\sin^2(\theta)}{2\sy^2}$}
\def \eqbgr  {$\gr = \bfrac{r}{\sx\sy} \exp\left(-a r^2\right) I_0(-b r^2)$}
\def \eqbgra {$a = \bfrac{\sx^2+\sy^2}{(2\sx\sy)^2}$}
\def \eqbgrb {$b = \bfrac{\sx^2-\sy^2}{(2\sx\sy)^2}$}

\def \eqcg   {$\g = \bfrac{1}{2\pi\sx\sy\sqrt{1-\p^2}} \exp\left(-\bfrac{1}{2(1-\p^2)}\left( \bfrac{x^2}{\sx^2} + \bfrac{y^2}{\sy^2} +\bfrac{2\p xy}{\sx\sy}\right) \right)$}
\def \eqcgt  {$\gt = \bfrac{1}{2ab}$}
\def \eqcgta {$a = 2\pi\sx\sy\sqrt{1-\p^2}$}
\def \eqcgtb {$b = \bfrac{1}{2(1-\p^2)}\left( \bfrac{\cos^2(\theta)}{\sx^2} + \bfrac{\sin^2(\theta)}{\sy^2} - \bfrac{2\p \cos(\theta)\sin(\theta)}{\sx\sy}\right)$}
\def \eqcgr  {$\gr = \bfrac{r}{\tsx\tsy} \exp\left(-a r^2\right) I_0(-b r^2)$}
\def \eqcgra {$a = \bfrac{\tsx^2+\tsy^2}{(2\tsx\tsy)^2}$}
\def \eqcgrb {$b = \bfrac{\tsx^2-\tsy^2}{(2\tsx\tsy)^2}$}
\def \eqcgrc {see Equations~(\ref{eqn:c-tsx}), (\ref{eqn:c-tsy}) for definition of $\tsx, \tsy$}

\def \eqdg   {$\g = \bfrac{1}{2\pi\s^2} \exp\left( - \bfrac{(x-\mx)^2 + (y-\my)^2}{2\s^2} \right)$}
\def \eqdgt  {$\gt = \bfrac{1}{\sqrt{2 \pi}} \phi(a)\left(1 + \bfrac{b\Phi(b)}{\phi(b)}\right)$}
\def \eqdgta {$a = \bfrac{1}{\s}\sqrt{\mx^2 + \my^2}$}
\def \eqdgtb {$b = \bfrac{1}{\s}(\mx\cos(\theta) + \my\sin(\theta))$}
\def \eqdgr  {$\gr = \bfrac{r}{\s^2} \exp\left(-\bfrac{r^2+a^2}{2\s^2}\right) I_0\left(\bfrac{ra}{\s^2}\right)$}
\def \eqdgra {$a = \sqrt{\mx^2 + \my^2}$}

\def \eqeg   {$\g = \bfrac{1}{2\pi\sx\sy} \exp\left(-\left(\bfrac{(x-\mx)^2}{2\sx^2} + \bfrac{(y-\my)^2}{2\sy^2} \right) \right)$}
\def \eqegt  {$\gt = \bfrac{1}{a \sqrt{2\pi \sx^2 \sy^2}} \phi(b) \left(1 + \bfrac{c\Phi(c)}{\phi(c)}\right)$}
\def \eqegta {$a = \bfrac{1}{\sx^2}\cos^2(\theta) + \bfrac{1}{\sy^2}\sin^2(\theta)$}
\def \eqegtb {$b = \sqrt{\bfrac{\mx^2}{\sx^2} + \bfrac{\my^2}{\sy^2}}$}
\def \eqegtc {$c = \bfrac{1}{\sqrt{a}} \left( \bfrac{\mx}{\sx^2}\cos(\theta) + \bfrac{\my}{\sy^2}\sin(\theta) \right)$}
\def \eqegr  {$\gr = a r \exp\left( -\bfrac{r^2(\sx^2 + \sy^2)}{4 \sx^2 \sy^2} \right) \left( I_0(br^2) I_0(cr) + 2 \sum_{k=1}^{\infty} I_k(br^2)I_{2k}(cr) \cos(2k \psi) \right)$}
\def \eqegra {$a = \bfrac{1}{\sx\sy} \exp \left( -\bfrac{\mx^2\sy^2 + \my^2\sx^2}{2\sx^2\sy^2} \right)$}
\def \eqegbb {$b = \bfrac{\sx^2 - \sy^2}{4\sx^2\sy^2}$}
\def \eqegrc {$c = \sqrt{ \left(\bfrac{\mx}{\sx^2}\right)^2 +  \left(\bfrac{\my}{\sy^2}\right)^2}$}
\def \eqegrd {$\psi = \tan^{-1} \left( \bfrac{\my \sx^2}{\mx \sy^2} \right)$}
 
\def \eqfg  {$\g = \bfrac{1}{2\pi\sx\sy\sqrt{1-\p^2}} \exp\left(-\bfrac{1}{2(1-\p^2)}\left( \bfrac{(x-\mx)^2}{\sx^2} + \bfrac{(y-\my)^2}{\sy^2} + \bfrac{2\p(x-\mx)(y-\my)}{\sx\sy}\right)\right)$}
\def \eqfgt  {$\gt = \bfrac{1}{a}\left(b + cd\Phi(d)\phi\left( \bfrac{c(\mx\sin(\theta) - \my\cos(\theta))}{\sqrt{a}}\right)\right)$}
\def \eqfgta {$a=c^2(\sy^2\cos^2(\theta) - \p\sx\sy\sin(2\theta) + \sx^2\sin^2(\theta))$}
\def \eqfgtb {$b=g(\mx,\my;0,0,\sx,\sy,\p)$}
\def \eqfgtc {$c=\bfrac{1}{\sx\sy\sqrt{1-\p^2}}$}
\def \eqfgtd{$d=\bfrac{c^2}{\sqrt{a}} (\mx\sy(\sy\cos(\theta) - \p\sx\sin(\theta)) + \my\sx(\sx\sin(\theta) - \p\sy\cos(\theta)))$}
\def \eqfgr  {$\gr = a r \exp\left( -\bfrac{r^2(\tsx^2 + \tsy^2)}{4 \tsx^2 \tsy^2} \right) \left( I_0(br^2) I_0(cr) + 2 \sum_{k=1}^{\infty} I_k(br^2)I_{2k}(cr) \cos(2k \psi) \right)$}
\def \eqfgra {$a = \bfrac{1}{\tsx\tsy} \exp \left( -\bfrac{\tmx^2\tsy^2 + \tmy^2\tsx^2}{2\tsx^2\tsy^2} \right)$}
\def \eqfgrb {$b = \bfrac{\tsx^2 - \tsy^2}{4\tsx^2\tsy^2}$}
\def \eqfgrc {$c = \sqrt{ \left(\bfrac{\tmx}{\tsx^2}\right)^2 +  \left(\bfrac{\tmy}{\tsy^2}\right)^2}$}
\def \eqfgrd {$\psi = \tan^{-1} \left( \bfrac{\tmy \tsx^2}{\tmx \tsy^2} \right)$}
\def \eqfgre {see Equations~(\ref{eqn:f-tmx})-(\ref{eqn:f-tsy}) for definition of $\tmx, \tmy, \tsx, \tsy$}
\begin{sidewaystable}
\begin{center}
\resizebox{0.9\textwidth}{!}{
\begin{tabular}{c|l|l|l|l}
 & case & bivariate normal, $\g$ & radial marginalization, $\gt$ & angular marginalization, $\gr$ \\
 &&&& \\
 \hline
 &&&& \\
%
(a) & \makecell[tl]{zero-mean [$\mx=0, \my=0$] \\ isotropic [$\s=\sx=\sy$, $\p = 0$]} & Equation~(\ref{eqn:a-g}) & Equation~(\ref{eqn:a-gt}) & Equation~(\ref{eqn:a-gr}), \emph{Rayleigh} \\
&& \eqag & \eqagt & \eqagr \\
&&&& \\
\hline
&&&& \\
%
(b) & \makecell[tl]{zero-mean [$\mx=0, \my=0$] \\ anisotropic [$\sx \neq \sy$] \\ diagonal [$\p = 0$]} & Equation~(\ref{eqn:b-g}) & Equation~(\ref{eqn:b-gt}) & Equation~(\ref{eqn:b-gr}) \\
&& \eqbg & \eqbgt & \eqbgr \\
&&&& \\
&&& \eqbgta & \eqbgra \qquad \eqbgrb \\
&&& \eqbgtb & \\
&&&& \\
\hline
&&&& \\
%
(c) & \makecell[tl]{zero-mean [$\mx=0, \my=0$] \\ anisotropic [$\sx \neq \sy$] \\ non-diagonal [$\p \neq 0$]} & Equation~(\ref{eqn:c-g}) & Equation~(\ref{eqn:c-gt}) & Equation~(\ref{eqn:c-gr}) \\
&& \eqcg & \eqcgt & \eqcgr \\
&&&& \\
&&& \eqcgta & \eqcgra \qquad \eqcgrb \\
&&& \eqcgtb & \eqcgrc \\
&&&& \\
\hline
&&&& \\
%
(d) & \makecell[tl]{non-zero-mean [$\mx \neq 0, \my \neq 0$] \\ isotropic [$\s=\sx=\sy$, $\p = 0$]} & Equation~(\ref{eqn:d-g}) & Equation~(\ref{eqn:d-gt}) & Equation~(\ref{eqn:d-gr}), \emph{Rician} \\
&& \eqdg & \eqdgt & \eqdgr \\
&&&& \\
&&& \eqdgta & \eqdgra \\
&&& \eqdgtb & \\
&&&& \\
\hline
&&&& \\
%
(e) & \makecell[tl]{non-zero-mean [$\mx \neq 0, \my \neq 0$] \\ anisotropic [$\sx \neq \sy$] \\ diagonal [$\p = 0$]} & Equation~(\ref{eqn:e-g}) & Equation~(\ref{eqn:e-gt}) & Equation~(\ref{eqn:e-gr}) \\
&& \eqeg & \eqegt & \eqegr \\
&&&& \\
&&& \eqegta & \eqegra \qquad \eqegbb \\
&&& \eqegtb & \\
&&& \eqegtc & \eqegrc \qquad \eqegrd \\
&&&& \\
\hline
&&&& \\
%
(f) & \makecell[tl]{non-zero-mean [$\mx \neq 0, \my \neq 0$] \\ anisotropic [$\sx \neq \sy$] \\ non-diagonal [$\p \neq 0$]}  & Equation~(\ref{eqn:f-g}) & Equation~(\ref{eqn:f-gt}) & Equation~(\ref{eqn:f-gr}) \\
&& \eqfg & \eqfgt & \eqfgr \\
&&&& \\
&&& \eqfgta & \eqfgra \qquad \eqfgrb \\
&&&& \\
&&& \eqfgtb & \eqfgrc \qquad \eqfgrd \\
&&&& \\
&&& \eqfgtc & \eqfgre \\
&&&& \\
&&& \eqfgtd & \\
&&&& \\
%
%
\end{tabular}
}
\end{center}
\label{tab:all}
\caption{The radial and angular marginalization of the bivariate normal distribution. In row (f) $g(x,y;\mx,\my,\sx,\sy,\p)$ denotes a bivariate normal with mean $\mx,\my$, variance $\sx,\sy$ and covariance $\p$ evaluated at $x,y$.}
\end{sidewaystable}
\newpage
\section{The Toolbox}
\label{sec:the-toolbox}

The full toolbox, implemented in Matlab and R, is available at \url{https://github.com/eacooper/RAMBiNo}. Below are snippets of the Matlab code for computing the analytic and numeric polar marginalizations.

\subsection{Analytic}

These Matlab code snippets compute  the polar marginalizations of arbitrary $2$-D normal distributions where \verb=pt= is $\gt$ and \verb=pr= is $\gr$. For each code snippet, the sampled angle ($\theta$) and radius ($r$) are denoted as \verb=t= and \verb=r=.

{\small
\noindent\blue{{\tt\% sampled angle and radius}}
\vspace{-1.0em}
\begin{verbatim}
t = [-180 : 0.1 : 179] * pi/180; % angle (radians)
r = [0 : 0.1 : 10]; % radius
\end{verbatim}
}

{\small
\noindent\blue{{\tt\% zero-mean, isotropic (Equation~(\ref{eqn:a-gt}) and~(\ref{eqn:a-gr}))}}
\vspace{-1.0em}
\begin{verbatim}
s  = 2; % variance

pt = 1/(2*pi)*ones(size(t));
pr = r/s^2 .* exp(-r.^2/(2*s^2));
\end{verbatim}
}

{\small
\noindent\blue{{\tt\% zero-mean, anisotropic, diagonal covariance (Equation~(\ref{eqn:b-gt}) and~(\ref{eqn:b-gr}))}}
\vspace{-1.0em}
\begin{verbatim}
sx = 3; % variance
sy = 2; % variance

a  = 2*pi*sx*sy;
b  = (cos(t).^2)/(2*sx^2) + (sin(t).^2)/(2*sy^2);
pt = 1./(2*a*b);

a  = (sy^2 + sx^2)/(2*sx*sy)^2;
b  = (sx^2 - sy^2)/(2*sx*sy)^2;
pr = r/(sx*sy) .* exp(-a*r.^2) .* besseli(0,-b*r.^2);
\end{verbatim}
}

{\small
\noindent\blue{{\tt\% zero-mean, anisotropic, non-diagonal covariance (Equation~(\ref{eqn:c-gt}) and~(\ref{eqn:c-gr}))}}
\vspace{-1.0em}
\begin{verbatim}
sx = 3; % variance
sy = 2; % variance
p  = 0.75; % covariance

a  = 2*pi*sx*sy*sqrt(1-p^2);
b  = 1/(2*(1-p^2)) * (cos(t).^2/(sx^2) + sin(t).^2/(sy^2) - 2*p*sin(t).*cos(t)/(sx*sy));
pt = 1./(2*a*b);

sxt = sqrt( (sx^2+sy^2)/2 + sqrt((sx^2+sy^2)^2/4 - (sx^2*sy^2 - p^2*sx^2*sy^2)) );
syt = sqrt( (sx^2+sy^2)/2 - sqrt((sx^2+sy^2)^2/4 - (sx^2*sy^2 - p^2*sx^2*sy^2)) );
a   = (sxt^2 + syt^2)/(2*sxt*syt)^2;
b   = (sxt^2 - syt^2)/(2*sxt*syt)^2;
pr  = r/(sxt*syt) .* exp(-a*r.^2) .* besseli(0,-b*r.^2);
\end{verbatim}
}

{\small
\noindent\blue{{\tt\% non-zero-mean, isotropic (Equation~(\ref{eqn:d-gt}) and~(\ref{eqn:d-gr}))}}
\vspace{-1.0em}
\begin{verbatim}
mx = 1.5; % mean
my = -1.5; % mean
s  = 2; % variance

a  = 1/s * sqrt(mx^2 + my^2);
b  = 1/s * (mx*cos(t) + my*sin(t));
pt = 1./(sqrt(2*pi)) .* normpdf(a) .* (1 + b.*normcdf(b)./normpdf(b));

a  = sqrt(mx^2 + my^2);
pr = r/s^2 .* exp(-(r.^2 + a^2)/(2*s^2)) .* besseli(0,(r*a)/s^2);
\end{verbatim}
}

{\small
\noindent\blue{{\tt\% non-zero-mean, anisotropic, diagonal covariance (Equation~(\ref{eqn:e-gt}) and~(\ref{eqn:e-gr}))}}
\vspace{-1.0em}
\begin{verbatim}
mx = 1.5; % mean
my = -1.5; % mean
sx = 3; % variance
sy = 2; % variance

a  = cos(t).^2/sx^2 + sin(t).^2/sy^2;
b  = sqrt(mx^2/sx^2 + my^2/sy^2);
c  = (cos(t)*mx/sx^2 + sin(t)*my/sy^2) ./ sqrt(a);
pt = 1./(a*sqrt(2*pi*sx^2*sy^2)) .* normpdf(b) .* (1 + c.*normcdf(c)./normpdf(c));

a   = 1/(sx*sy) * exp(-(mx^2*sy^2 + my^2*sx^2)/(2*sx^2*sy^2));
b   = (sx^2-sy^2)/(4*sx^2*sy^2);
c   = sqrt((mx/sx^2)^2 + (my/sy^2)^2);
psi = atan2( (my*sx^2), (mx*sy^2) );
d   = zeros( size(r) );
for k = 1 : 100 % truncated series
    d = d + (besseli(k,b*r.^2) .* besseli(2*k,c*r) * cos(2*k*psi));
end
pr = a*r .* exp(-(r.^2*(sx^2+sy^2))/(4*sx^2*sy^2)) .* \\
     (besseli(0,b*r.^2) .* besseli(0,c*r) + 2*d);
\end{verbatim}
}

{\small
\noindent\blue{{\tt\% non-zero-mean, anisotropic, non-diagonal covariance (Equation~(\ref{eqn:f-gt}) and~(\ref{eqn:f-gr}))}}
\vspace{-1.0em}
\begin{verbatim}
mx = 1.5; % mean 
my = -1.5; % mean
sx = 3; % variance
sy = 2; % variance
p  = 0.75; % covariance

c  = 1 / (sx*sy*sqrt(1-p^2));
a  = c^2 * (sy^2*cos(t).^2 - p*sx*sy*sin(2*t) + sx^2*sin(t).^2);
b  = generateN(mx,my,0,0,sx,sy,p);
d  = (c^2./sqrt(a)) .* (mx*sy*(sy*cos(t) - p*sx*sin(t)) + my*sx*(sx*sin(t) - p*sy*cos(t)));
pt = 1./a .* (b + c*d.*normcdf(d).*normpdf((c*(mx*sin(t) - my*cos(t)))./sqrt(a)) );

w   = 1/2*atan2((2*p*sx*sy), (sx^2-sy^2));
mxt = mx*cos(w) + my*sin(w);
myt = -mx*sin(w) + my*cos(w);
sxt = sqrt( (sx^2+sy^2)/2 + sqrt((sx^2+sy^2)^2/4 - (sx^2*sy^2 - p^2*sx^2*sy^2)) );
syt = sqrt( (sx^2+sy^2)/2 - sqrt((sx^2+sy^2)^2/4 - (sx^2*sy^2 - p^2*sx^2*sy^2)) );
a   = 1/(sxt*syt) * exp(-(mxt^2*syt^2 + myt^2*sxt^2)/(2*sxt^2*syt^2));
b   = (sxt^2-syt^2)/(4*sxt^2*syt^2);
c   = sqrt((mxt/sxt^2)^2 + (myt/syt^2)^2);
psi = atan2( (myt*sxt^2), (mxt*syt^2) );
d   = zeros( size(r) );
for k = 1 : 100 % truncated series
    d = d + (besseli(k,b*r.^2) .* besseli(2*k,c*r) * cos(2*k*psi));
end
pr = a*r .* exp(-(r.^2*(sxt^2+syt^2))/(4*sxt^2*syt^2)) .* \\
     (besseli(0,b*r.^2) .* besseli(0,c*r) + 2*d);
\end{verbatim}
}

\subsection{Numeric}

These Matlab functions numerically compute the polar marginalizations of arbitrary $2$-D normal distributions.

{\small
\noindent \blue{{\tt \% input: a 2-D normal with mean (mx,my), variance (sx,sy), and covariance (p)}}

\noindent \blue{{\tt \% output: angle (t) and distribution over angle (pt); and }}

\noindent \blue{{\tt \%         radius (r) and distribution over radius (pr)}}

\noindent \blue{{\tt \% example:}}

\noindent \blue{{\tt\%   [t,pt,r,pr]=marginalize(0,0,2,2,0); \% zero-mean, isotropic}}

\noindent \blue{{\tt\%   [t,pt,r,pr]=marginalize(0,0,3,2,0); \% zero-mean, anisotropic, diagonal}}

\noindent \blue{{\tt\%   [t,pt,r,pr]=marginalize(0,0,3,2,0.75); \% zero-mean, anisotropic, non-diagonal}}

\noindent \blue{{\tt\%   [t,pt,r,pr]=marginalize(1.5,-1.5,2,2,0); \% non-zero-mean, isotropic}}

\noindent \blue{{\tt\%   [t,pt,r,pr]=marginalize(1.5,-1.5,3,2,0); \% non-zero-mean, anisotropic, diagonal}}

\noindent \blue{{\tt\%   [t,pt,r,pr]=marginalize(1.5,-1.5,3,2,0.75); \% non-zero-mean, anisotropic, diagonal}}
}
\vspace{-1.0em}

{\small
\begin{verbatim}
function[t,pt,r,pr] = marginalize( mx, my, sx, sy, p )
[t,pt] = marginalizeR( mx, my, sx, sy, p ); % marginalize over radius
[r,pr] = marginalizeT( mx, my, sx, sy, p ); % marginalize over angle
\end{verbatim}

\noindent \blue{{\tt \% generate a bivariate normal distribution}}
\vspace{-1.0em}
\begin{verbatim}
function[N] = generateN( x, y, mx, my, sx, sy, p )
N = 1/(2*pi*sx*sy*sqrt(1-p^2)) * \\
    exp( -(1/(2*(1-p^2))*((x-mx).^2/(sx^2) + (y-my).^2/(sy^2) - 2*p*(x-mx).*(y-my)/(sx*sy))) );
\end{verbatim}

\noindent \blue{{\tt \%  marginalization over radius (returns angle t and p(t))}}
\vspace{-1.0em}
\begin{verbatim}
%  marginalization over radius (returns angle T and p(t))
function[T,pt] = numeric_pt( mx, my, sx, sy, p, T, R )
pt = zeros(1,length(T));
C  = 2*pi*R; % circumference
k = 1;
for t = T
    x = R*cos(t);
    y = R*sin(t);
    pt(k) = sum( C .* generateN( x, y, mx, my, sx, sy, p ) );
    k = k + 1;
end
pt = length(pt) * 1/(2*pi) * pt/sum(pt); % mean of distribution should be 1/(2*pi)
\end{verbatim}

\noindent \blue{{\tt \% marginalization over angle (returns radius r and p(r))}}
\vspace{-1.0em}
\begin{verbatim}
function[R,pr] = numeric_pr( mx, my, sx, sy, p, T, R )
pr = zeros(1,length(R));
C  = 2*pi*R; % circumference
k = 1;
for r = R
    x = r*cos(T);
    y = r*sin(T);
    pr(k) = mean( C(k) .* generateN( x, y, mx, my, sx, sy, p ) );
    k = k + 1;
end
\end{verbatim}
}

\bibliography{main}

\end{document}